\documentclass[a4paper,12pt]{article}

\usepackage[latin1]{inputenc}
\usepackage[top=2.5cm, left=2.5cm, right=2cm, bottom=2.5cm]{geometry}
\usepackage{graphicx}
\usepackage[centertags]{amsmath}
\usepackage{amsfonts}
\usepackage{amssymb}
\usepackage{amsthm}
\usepackage{newlfont}

\begin{document}

\begin{center}

{\Large Constraint Analysis of Two-Dimensional Quadratic Gravity from $BF$ Theory}

\vspace{1cm}

C. E. Valc\'arcel\footnote{valcarcel.flores@gmail.com}

\vspace{.5cm}

\emph{Centro de Matem\'atica, Computa\c{c}\~ao e Cogni\c{c}\~ao},

\emph{Universidade Federal do ABC, CEP 09210-170, Santo Andr\'e, SP, Brazil}.

\end{center}

\vspace{.25cm}

\begin{abstract}

Quadratic gravity in two dimensions can be formulated as a Background Field
($BF$) theory plus an interaction term which is polynomial in both, the gauge
and Background fields. This formulation is similar to the one given by Freidel and
Starodubtsev to obtain MacDowell-Mansouri gravity in four dimensions.

In this article we use the Dirac's Hamiltonian formalism to analyze the  constraint
structure of the two-dimensional Polynomial $BF$ action. After we obtain the constraints
of the theory, we proceed with the Batalin-Fradkin-Vilkovisky procedure to obtain the
transition amplitude. We also compare our results with the ones obtained from generalized
dilaton gravity.

\vspace{.5cm}


\end{abstract}

\section{Introduction}

In their seminal work \cite{MacDowell}, MacDowell and Mansouri translate the Palatini action of General Relativity into a Yang-Mills like gauge theory of gravity. In fact, the MacDowell-Mansouri and Palatini actions are equivalent up to topological terms. These topological terms can render finite Noether charges and recover a regularized action for $AdS$ gravity \cite{Olea}. Recently, it has been shown \cite{Smolin,Freidel} that MacDowell-Mansouri gravity can be formulated as a Background Field ($BF$) theory plus a break invariance term. This formulation relates gravity with topological field theories. Furthermore, $BF$ models of gravity are good laboratories for the study of spin foam quantization \cite{Perez} and loop quantization \cite{Ashtekar}.

Despite all efforts, there is still no satisfactory four-dimensional quantum theory gravity. Then, we can find in the study of lower dimensional gravity \cite{Brown} good models to understand some quantum properties of the gravitational field. However, lower dimensional models have some trivialities. For example, in two dimensions, the Einstein tensor is identically zero, and leads to trivial equations of motion. One way to deal with this problem was proposed by Jackiw and Teitelboim \cite{Jackiw-Teitelboim}. They proposed the introduction of a dilaton field, such that, the equation of motion $R=2\Lambda$, where $R$ is the curvature scalar and $\Lambda$ the cosmological constant, holds. It is important to notice that Jackiw-Teitelboim gravity is also a class of a more general family of two-dimensional dilaton theories \cite{Grumiller}. In \cite{Isler,Chamseddine}, it was shown that Jackiw-Teitelboim gravity can be written as a $BF$ gauge theory under the group $SO\left(2,1\right)$.

Recently, in \cite{Paszko}, a $BF$ theory for two and three-dimensional gravity has been build in analogy with the construction of MacDowell-Mansouri. The key ingredient is the introduction of an interaction term, polynomial in both: the background and gauge field. For two-dimensional gravity, this interaction breaks the $SO\left(3\right)$ invariance of the $BF$ theory and, as a consequence, a quadratic term in the curvature appears. Due to the presence of this term, the Einstein tensor is no longer zero. Furthermore, it is also known $f\left(R\right)$ models allow the study of inflationary models, see for example \cite{Felice}.

In this work, our objective is to use the Dirac's Hamiltonian formalism \cite{Dirac} to analyze the constraint structure of the $BF$ model equivalent to quadratic gravity. This analysis allows us to obtain the symmetries of the model and is the first step to canonical and loop quantization. Until now, we have no found a constraint analysis of the Polynomial $BF$ action. However, a canonical analysis for the $BF$ action equivalent to Jackiw-Teitelboim can be found in \cite{Piguet} using the Dirac's formalism, and in \cite{BF-HJ} under the Hamilton-Jacobi formalism \cite{HJ}.

In the next section, we will review the Polynomial $BF$ action in two dimensions and show its equivalence with quadratic gravity. In section $3$, we will use the Dirac's Hamiltonian programme to obtain the first-class constraints that are generators of the symmetries of the Polynomial $BF$ action. In section $4$, we will also use the Batalin-Fradkin-Vilkovisky \cite{BFV} formalism to obtain the path integral transition amplitude of the model. In section $5$ we will briefly resume a class of generalized dilaton theories which resemble quadratic gravity and compare with the ones given by the Polynomial $BF$ action. Finally, in section $6$ we will discuss our results.

\section{Two-dimensional Polynomial $BF$ Action}

Let us review the equivalence between the two-dimensional Polynomial $BF$ action and quadratic gravity, as shown in
\cite{Paszko}. The basic idea is to build a $SO\left(3\right)$ invariant $BF$ action with an interaction term that will breaks this invariance. The Polynomial $BF$ action is given by
\begin{eqnarray}
S=\int_{\mathcal{M}}\: tr\left[-\frac{1}{2}B  F\left(A\right)+\kappa^{2}\left(PB\right)\left(PB\right)\left(QA\right)\wedge A\right],\label{BF01}
\end{eqnarray}
where $\kappa$ is a constant, $B$ is a $0-$form $B=\frac{1}{2}B^{IJ}M_{IJ}$, the $1-$form gauge field is $A=\frac{1}{2}A_{\mu}^{IJ}M_{IJ}dx^{\mu}$ and $F=dA+A\wedge A$ is its respective field strength. The generators of the algebra $SO\left(3\right)$ are $M_{IJ}=-M_{JI}$ (with $I=0,1,2$) and metric $\eta_{IJ}=diag\left(+,+,+\right)$. $P$ and $Q$ are projection operators defined by
\begin{eqnarray}
P^{IJ,KL} & \equiv & \frac{1}{2}\epsilon^{IJ2}\epsilon^{KL2}.\label{BF02a}\\
Q_{IJ,KL} & \equiv & \frac{1}{2}\epsilon_{IJ}^{\;\;\;\;\; M}\epsilon_{KL}^{\;\;\;\;\; N}\epsilon_{MN2}.\label{BF02b}
\end{eqnarray}
where $\epsilon^{IJK}$ is the Levi-Civita symbol and, by convention we have $\epsilon^{012}=1$.

We can decompose the generators $M_{IJ}$ into the generators $M_{ab}$ $\left(a=0,1\right)$ and the generators $P_{a}\equiv M_{a2}$. It is easy to see that the operator (\ref{BF02a}) does not project any $M_{a2}$, while leave invariant the generators $M_{ab}$
\begin{eqnarray}
\left(PM\right)_{ab}=M_{ab}, & \; & \left(PM\right)_{a2}=0,\label{BF03a}
\end{eqnarray}
and the operator (\ref{BF02b}) does not project any generator $M_{ab}$ and interchange the generators $M_{a2}$ \begin{eqnarray}
\left(QM\right)_{a2}=\epsilon_{a}^{\;\; b}M_{b2} & \; & \left(QM\right)_{ab}=0.\label{BF03b}
\end{eqnarray}
Explicitly, the two-dimensional Polynomial $BF$ action reads
\begin{eqnarray}
S=\int d^{2}x\:\varepsilon^{\mu\nu}\left[\frac{1}{4}B_{KL}F_{\mu\nu}^{KL}
+\frac{\kappa^{2}}{2}\left(B^{ab}B_{ab}\right)\epsilon_{mn}A_{\mu}^{m}A_{\nu}^{n}\right],\label{BF03c}
\end{eqnarray}
where $\varepsilon^{\mu\nu}$ is the Levi-Civita symbol in the curved space. Therefore, the interaction term in (\ref{BF01}) breaks the $SO\left(3\right)$ invariance.

The gauge field can be expressed in terms of the spin connection $\omega_{\mu}^{ab}=\epsilon^{ab}\omega_{\mu}$ and the zweibein $e_{\mu}^{a}$ as
\begin{equation}
A_{\mu}=\frac{1}{2}\omega_{\mu}\epsilon^{ab}M_{ab}+\frac{1}{l}e^{a}P_{a},\label{BF04}
\end{equation}
where $l$ is a parameter. Having an explicit form for the gauge field, we can decompose the components of the field strength
\begin{eqnarray}
F_{\mu\nu}^{a2} & = & \frac{1}{l}T_{\mu\nu}^{a},\label{BF05a}\\
F_{\mu\nu}^{ab} & = & R_{\mu\nu}^{ab}-\frac{1}{l^{2}}\left(e_{\mu}^{a}e_{\nu}^{b}-e_{\nu}^{a}e_{\mu}^{b}\right),\label{BF05b}
\end{eqnarray}
where $T_{\mu\nu}^{a}$ and $R_{\mu\nu}^{ab}$ are the components of the Torsion $T=de+\omega\wedge e$ and Curvature forms
$R=d\omega+\omega\wedge\omega$ respectively. In two-dimensions we have
\begin{eqnarray}
T_{\mu\nu}^{a} & = & \left(\partial_{\mu}e_{\nu}^{a}+\omega_{\mu}\epsilon_{\; b}^{a}e_{\nu}^{b}\right)-\left(\partial_{\nu}e_{\mu}^{a}+\omega_{\nu}\epsilon_{\; b}^{a}e_{\mu}^{b}\right),\label{BF06a}\\
R_{\mu\nu}^{ab} & = & \epsilon^{ab}\left(\partial_{\mu}\omega_{\nu}-\partial_{\nu}\omega_{\mu}\right).\label{BF06b}
\end{eqnarray}
Replacing (\ref{BF04}) and (\ref{BF05a}),(\ref{BF05b}) into the Polynomial $BF$ action, we obtain
\begin{equation}
S=\int d^{2}x\:\left[\frac{1}{2}B\left(\frac{1}{2}\varepsilon^{\mu\nu}\epsilon_{ab}R_{\mu\nu}^{ab}-\frac{2}{l^{2}}e\right)+\frac{1}{2l}\varepsilon^{\mu\nu}B_{a}T_{\mu\nu}^{a}+\frac{2\kappa^{2}}{l^{2}}B^{2}e\right],\label{BF07}
\end{equation}
where $e=\det\left(e_{\mu}^{a}\right)$ is the determinant of the zweibein, and $B_{a}\equiv B_{a2}$, $B_{ab}=\epsilon_{ab}B$. Note that the variation of the fields $B_{a}$ and $B_{ab}$ give the following equations
\begin{eqnarray}
0 & = & \varepsilon^{\mu\nu}T_{\mu\nu}^{a},\label{BF08a}\\
0 & = & \frac{1}{4}\varepsilon^{\mu\nu}\epsilon_{ab}R_{\mu\nu}^{ab}-\frac{1}{l^{2}}e+\frac{4\kappa^{2}e}{l^{2}}B.\label{BF08b}
\end{eqnarray}

Equation (\ref{BF08a}) is the torsion-free condition, since we are working in two dimensions, this equation becomes
\begin{equation}
0=\partial_{0}e_{1}^{a}-\partial_{1}e_{0}^{a}+\epsilon_{\; b}^{a}\omega_{0}e_{1}^{b}-\epsilon_{\; b}^{a}\omega_{1}e_{0}^{b}.\label{BF09}
\end{equation}
Note that, as long as the zweibein is invertible, it is possible to solve (\ref{BF09}) for the connection in terms of the tetrad. Therefore, the variables $\left(e_{\mu}^{a},\omega_{\mu}\right)$ are no longer independents. In fact, now they relate the curvature $R_{\mu\nu}^{ab}$ with the Riemann tensor $R_{\mu\nu}^{\:\:\alpha\beta}=R_{\mu\nu}^{ab}e_{a}^{\alpha}e_{b}^{\beta}$. From now on, we consider that the torsion-free condition is always satisfied.

Now, let us analyze equation (\ref{BF08b}). Notice that, if $\kappa=0$, i.e., there is no interaction term. The $B$ field remains arbitrary. Furthermore, by choosing the cosmological constant $\Lambda=1/l^{2}$, the action (\ref{BF07}) becomes the Jackiw-Teitelboim action
\begin{equation}
S=\frac{1}{2}\int d^{2}x\:\sqrt{g}B\left(R-2\Lambda\right),\label{BF10}
\end{equation}
where $R$ is the curvature scalar $R=g^{\mu\nu}R_{\:\mu\alpha\nu}^{\alpha}$. The equation of motion (\ref{BF08b}) becomes $R=2\Lambda$, and we can identity $B$ as a dilaton field. This procedure is similar to the one given by \cite{Chamseddine}.

On the other hand, for $\kappa$ different than zero, equation (\ref{BF08b}) relates the $B$ field with the curvature and the parameter $l^{2}$ as
\begin{eqnarray}
B & = & -\frac{l^{2}}{8\kappa^{2}}\left(R-\frac{2}{l^{2}}\right).\label{BF11}
\end{eqnarray}
Replacing this condition on the action and, for $2\Lambda=1/l^{2}$, we obtain the following quadratic gravity action
\begin{eqnarray}
S & = & \frac{1}{8\kappa^{2}}\int d^{2}x\:\sqrt{g}\left(R-2\Lambda\right)-\frac{1}{64\Lambda\kappa^{2}}\int d^{2}x\:\sqrt{g}R^{2}.\label{BF12}
\end{eqnarray}
Finally, we can choose $\kappa^{2}=2\pi G$ in order to have usual coefficient in front of the Einstein-Hilbert
action.  Therefore, we can say that the two-dimensional Polynomial $BF$ action (\ref{BF01}) is equivalent to quadratic gravity.

\section{Hamiltonian Constraint Analysis}

In order to proceed with the constraint analysis, it is necessary to introduce a foliation of in space-time: $\mathcal{M}=\mathbb{R}\times M_{1}$. Here $M_{1}$ represents a Cauchy space at constant time $x^{0}=t=cte$. After eliminating some boundary terms, it is possible to write the action (\ref{BF07}) as
\begin{equation}
S=\int d^{2}x\:\left[B\partial_{0}\omega+\frac{1}{l}B_{a}\partial_{0}e_{1}^{a}+\omega_{0}D_{1}B+\frac{1}{l}e_{0}^{a}\left(D_{1}B_{a}+\frac{2\kappa^{2}}{l}\epsilon_{ab}e_{1}^{b}B^{2}\right)\right].\label{D01}
\end{equation}
Note that we had dropped the notation with the gauge field $A^{I}$ and we are using its components $\left(e_{\mu}^{a},\omega_{\mu}^{ab}\right)$ instead. Furthermore, we have written $\omega\equiv\omega_{1}$. The covariant derivatives are given by
\begin{eqnarray}
D_{1}B_{a} & = & \partial_{1}B_{a}+\epsilon_{a}^{\; c}B_{c}\omega-\frac{1}{l}\epsilon_{ac}e_{1}^{c}B.\label{D02a}\\
D_{1}B & = & \partial_{1}B+\frac{1}{l}\epsilon_{\; b}^{a}e_{1}^{b}B_{a}.\label{D02b}
\end{eqnarray}
From (\ref{D01}) we have that $B$ and $B_{a}$ are proportional to the canonical momenta of $\omega$ and $e_{1}^{a}$, respectively. Therefore, we can write
\begin{equation}
\left\{ e_{1}^{a}\left(x\right),B_{b}\left(y\right)\right\} =l\delta_{b}^{a}\delta\left(x-y\right),\:\left\{ \omega\left(x\right),B\left(y\right)\right\} =\delta\left(x-y\right).\label{D03}
\end{equation}

On the other hand, the expressions for the canonical momenta of the variables $\left(e_{0}^{a},\omega_{0}\right)$
 represent primary constraints
\begin{equation}
\phi_{a}\equiv\pi_{a}^{0}\approx0,\;\phi\equiv\Pi^{0}\approx0,\label{D04}
\end{equation}
where $\approx$ means weak equality. Having all the expressions for the canonical momenta, we build the canonical Hamiltonian
\begin{equation}
H_{0}=-\int dx\:\left[\omega_{0}D_{1}B+\frac{1}{l}e_{0}^{a}\left(D_{1}B_{a}+\frac{2\kappa^{2}}{l}\epsilon_{ab}e_{1}^{b}B^{2}\right)\right].\label{D05}
\end{equation}
The primary Hamiltonian is defined as $\mathcal H_P \equiv H_0 + \mu \phi + \mu^a \phi_a$, where $\mu,\mu^a$ are Lagrange multipliers. Now, the consistency condition states that the constraints (\ref{D04}) must be preserved in time. This condition generates two secondary constraints:
\begin{eqnarray}
\dot{\phi}=\{\phi, H_P\}=0 & \rightarrow & \mathcal{G}\equiv D_{1}B\approx0,\label{D06a}\\
\dot{\phi}_{a}=\{\phi, H_P\}=0 & \rightarrow & \mathcal{G}_{a}\equiv D_{1}B_{a}+\frac{2\kappa^{2}}{l}\epsilon_{ab}e_{1}^{b}B^{2}\approx0.\label{D06b}
\end{eqnarray}
The canonical Hamiltonian (\ref{D05}) is now a linear combination of the secondary constraints $\mathcal{H}_{0}=-\omega_{0}\mathcal{G}-\frac{1}{l}e_{0}^{a}\mathcal{G}_{a}$. This is the case for generally covariant theories and enforce the fact that the variables $\left(e_{0}^{a},\omega_{0}^{ab}\right)$ act as Lagrange
multipliers. The secondary constraints satisfy the following algebra
\begin{eqnarray}
\left\{ \mathcal{G}\left(x\right),\mathcal{G}\left(y\right)\right\}  & = & 0,\label{D07a}\\
\left\{ \mathcal{G}\left(x\right),\mathcal{G}_{a}\left(y\right)\right\}  & = & -\epsilon_{a}^{\;\; b}\mathcal{G}_{b}\left(x\right)\delta\left(x-y\right),\label{D07b}\\
\left\{ \mathcal{G}_{a}\left(x\right),\mathcal{G}_{b}\left(y\right)\right\}  & = & -\epsilon_{ab}\left(1-4\kappa^{2}B\right)\mathcal{G}\left(x\right)\delta\left(x-y\right).\label{D07c}
\end{eqnarray}
Also notice the dependence on the Background field component $B=B\left(x\right)$ in equation (\ref{D07c}).

The consistency condition for the secondary constraints reads
\begin{eqnarray}
\dot{\mathcal G}=\{ \mathcal G, H_P\} = \{ \mathcal G, H_0\} + \mu\{\mathcal G,\phi\} + \mu^a \{ \mathcal G,\phi_a\},\label{D06a}\\
\dot{\mathcal G}_{a}=\{\mathcal G_a, H_P\} =\{ \mathcal G_a, H_0\} + \mu\{ \mathcal G_a,\phi \} + \mu^b \{ \mathcal G_a,\phi_b\}.\label{D06b}
\end{eqnarray}
Once that, the canonical Hamiltonian is a combination of the secondary constraints, the consistency condition is satisfied, $\dot{\mathcal G} \approx 0, \dot{\mathcal G}_a \approx 0$, and no tertiary constraints are found. Furthermore, the Lagrangian multipliers $\mu,\mu^a$ remains undetermined.

As we mention in the previous section, if we set $\kappa=0$ we are no longer in the case of quadratic gravity. However, as $\kappa\rightarrow0$, we approach to the $SO\left(3\right)$ algebra (see equation (\ref{AP02}) in the Appendix). Furthermore, the sets $\left(\phi,\phi_{a}\right)$ and $\left(\mathcal{G},\mathcal{G}_{a}\right)$ commute, since the first one does not depend on the temporal components of the zweibein and the spin-connection. This means that all primary and secondary constraints are first-class.

It is usual to work with ``smeared out'' constraints, which, in our case are
\begin{eqnarray}
\mathcal{G}\left(\zeta\right)\equiv\int dx\:\zeta\left(x\right)\mathcal{G}\left(x\right), & \; & \mathcal{G}_{a}\left(\zeta^{a}\right)\equiv\int dx\:\zeta^{a}\left(x\right)\mathcal{G}_{a}\left(x\right),\label{D08a}\\
\mathcal{\widetilde{G}}\left(\lambda\right)\equiv\int dx\:\lambda\left(x\right)\phi\left(x\right), & \; & \mathcal{\mathcal{\widetilde{G}}}_{a}\left(\lambda^{a}\right)\equiv\int dx\:\lambda^{a}\left(x\right)\phi_{a}\left(x\right).\label{D08b}
\end{eqnarray}
which depend on the parameters $\left(\eta,\lambda,\eta^{a},\lambda^{a}\right)$. Smeared constraints are used to avoid distributions and are usually found in loop quantization language \cite{Gambini}.

The smeared primary secondary constraints state that the variables $\left(e_{0}^{a},\omega_{0}\right)$ are arbitrary
\begin{eqnarray}
\left\{ e_{0}^{a},\mathcal{\mathcal{\widetilde{G}}}_{b}\left(\lambda^{b}\right)\right\} = \lambda^{a}, \:\left\{ \omega_{0},\mathcal{\mathcal{\widetilde{G}}}\left(\lambda\right)\right\} =\lambda.\label{D09}
\end{eqnarray}

On the other hand, the smeared secondary constraints acts as generators of gauge and shift transformations for the dynamical fields. This can be checked by defining the sum
\begin{equation}
G_{en}\left(\zeta,\zeta^{a}\right)\equiv\mathcal{G}\left(\zeta\right)+\mathcal{G}_{a}\left(\zeta^{a}\right),\label{D10}
\end{equation}
and see that for the $\left(e_{1}^{a},\omega\right)$ we have
\begin{eqnarray}
\left\{ e_{1}^{a},G_{en}\right\}  & = & l D_{1}\zeta^{a},\label{D11a}\\
\left\{ \omega,G_{en}\right\}  & = & -D_{1}\zeta+\frac{4\kappa^{2}}{l}\epsilon_{bc}\zeta^{b}e_{1}^{c}B,\label{D11b}
\end{eqnarray}
while for $\left(B_{a},B\right)$ we obtain
\begin{eqnarray}
\left\{ B_{a},G_{en}\right\}  & = & \epsilon_{a}^{\: b}B_{b}\zeta-\epsilon_{ab}\zeta^{b}B+2\kappa^{2}\epsilon_{ab}\zeta^{b}B^{2},\label{D11c}\\
\left\{ B,G_{en}\right\}  & = & -\epsilon_{b}^{\: c}B_{c}\zeta^{b}.\label{D11d}
\end{eqnarray}
The phase-space of the two-dimensional Polynomial $BF$ action has twelve dimensions, related to the variables $\left(e_{\mu}^{a},\omega_{\mu}\right)$ and their respective canonical momenta. We also have six first-class constraints: $\left(\phi,\phi_{a},\mathcal{G},\mathcal{G}_{a}\right)$ and no second-class constraints. Consequently, we have zero degrees of freedom, in perfect agreement with the topological nature of the model.

\section{Batalin-Fradkin-Vilkovisky Path Integral Formulation}

The presence of first-class constraints is related to gauge symmetries in the theory. One way to remove the unobservable gauge freedom is to introduce accessible gauge fixing constraints. Fixing the gauge freedom completely means that we introduce one gauge fixing constraint for each first-class constraint such that the new complete set of constraints is second-class. After isolate the true degrees of freedom, we can proceed with the canonical quantization. Another approach was given by Batalin, Fradkin and Vilkovisky (BFV) \cite{BFV}, which incorporates the notion of constraints in the Path Integral Quantization.

To begin with the BFV quantization, let us denote $G_{A}=\left(\phi,\phi_{a},\mathcal{G},\mathcal{G}_{a}\right)$ the collection of all primary and secondary constraints. The sub-index $A$ goes from $1$ to $4$. Having a system full of first-class constraints, they satisfy $\left\{ G_{A},G_{B}\right\} =U_{AB}^{\;\;\;\;\; C}G_{C}$ and $\left\{ H_{0},G_{A}\right\} =V_{A}^{B}G_{B}$. In general, the structure coefficients $U_{AB}^{\;\;\;\;\; C}$ and $V_{A}^{B}$ are functions of the canonical variables.  For constant $U_{AB}^{\; C}$l, as in Jackiw-Teitelboim gravity, the system close a Lie algebra.

From the previous section, we have
\begin{equation}
U_{34}^{\;\;\; 4}=-\epsilon_{a}^{\; b},\: U_{44}^{\;\;\; 3}=-\epsilon_{ab}\left(1-4\kappa^{2}B\right).\label{Q01}
\end{equation}
We notice that $U_{44}^{\;\;\; 3}$ depends on the Background field component $B$, nonetheless, we still have an involutive algebra. The non-zero $V_{A}^{B}$ are
\begin{equation}
V_{1}^{3}=-1,\: V_{2}^{4}=-\frac{1}{l}\delta^b_a,\: V_{3}^{4}=-\frac{1}{l}\epsilon_{a}^{\; b}e_{0}^{a},\: V_{4}^{3}=-\frac{1}{l}\epsilon_{ab}e_{0}^{b}\left(1-4\kappa^{2}B\right),\: V_{4}^{4}=\epsilon_{a}^{\; b}\omega_{0}.\label{Q02}
\end{equation}
The dependence of $V_{A}^{B}$ with the canonical variables is also evident.

Now, let us introduce the vector of ghost $\eta^{A}=\left(P,P^{a},c,c^{a}\right)$ (with ghost number equal to $1$), and its respective canonical momenta $\overline{\mathcal{P}}_{A}=\left(\overline{c},\overline{c}_{a},\overline{P},\overline{P}_{a}\right)$ (with ghost number equal to $-1$). All of these ghost satisfy the following graded brackets
\begin{eqnarray}
\left\{ P,\overline{c}\right\} =-\delta\left(x-y\right), & \: & \left\{ P^{a},\overline{c}_{b}\right\} =-\delta_{b}^{a}\delta\left(x-y\right),\label{Q03a}\\
\left\{ c,\overline{P}\right\} =-\delta\left(x-y\right), & \: & \left\{ c^{a},\overline{P}_{b}\right\} =-\delta_{b}^{a}\delta\left(x-y\right).\label{Q03b}
\end{eqnarray}

We can now introduce the BRST charge $\Omega$, defined to be nilpotent $\Omega^{2}=\frac{1}{2}\left\{ \Omega,\Omega\right\} =0$. An extensive explanation on the construction of the BRST charge can be found \cite{Henneaux}. In this work we follow the notation of \cite{Rothe}. For involutive algebras, the BRST charge is given by $\Omega=\eta^{A}G_{A}+\frac{1}{2}\overline{\mathcal{P}}_{C}U_{AB}^{\;\;\;\;\; C}\eta^{A}\eta^{B}$. In our case we have
\begin{equation}
\Omega=P\phi+P^{a}\phi_{a}+c\mathcal{G}+c^{a}\mathcal{G}_{a}-\frac{1}{2}\epsilon_{ab}\left(1-4\kappa^{2}B\right)\overline{P}c^{a}c^{b}+\epsilon_{\; b}^{a}\overline{P}_{a}cc^{b}.\label{Q04}
\end{equation}

We can build a BRST invariant Hamiltonian $\mathcal{H}_{B}$, which satisfy $\left\{ \mathcal{H}_{B},\Omega\right\} =0$. This Hamiltonian is build with the ghost, their momenta and the structure coefficients $V_{A}^{B}$. It has the form
$\mathcal{H}_{B}=\mathcal{H}_{0}+\eta^{A}V_{A}^{B}\overline{\mathcal{P}}_{B}$. By replacing all the values from
(\ref{Q02}), we obtain
\begin{equation}
\mathcal{H}_{B}=-\omega_{0}\mathcal{G}-\frac{1}{l}e_{0}^{a}\mathcal{G}_{a}-P\overline{P}-\frac{1}{l}P^{a}\overline{P}_{a}-\frac{1}{l}\epsilon_{a}^{\; b}e_{0}^{a}c\overline{P}_{b}-\frac{1}{l}\epsilon_{ab}e_{0}^{b}\left(1-4\kappa^{2}B\right)c^{a}\overline{P}+\epsilon_{a}^{\; b}\omega_{0}c^{a}\overline{P}_{b}.\label{Q05}
\end{equation}
There is, however, an indeterminacy in this construction, since we can add exact charges to this Hamiltonian without modifying their BRST invariance. In order to get rid of this indeterminacy, we define the unitarizing Hamiltonian
\begin{equation}
\mathcal{H}_{U}=\mathcal{H}_{B}+\left\{ \Psi,\Omega\right\} ,\label{Q06}
\end{equation}
where $\Psi$ is called gauge-fixing fermion. The corresponding Lagrangian to (\ref{Q06}), which also consider the dynamic of the ghost fields, is given by $\mathcal{L}_{U}=\dot{q}_{i}p^{i}+\dot{\eta}^{A}\overline{\mathcal{P}}_{A}-\mathcal{H}_{U}$. Explicitly, for quadratic gravity we have
\begin{equation}
\mathcal{L}_{U}=\dot{e}_{0}^{a}\pi_{a}^{0}+\dot{e}_{1}^{a}B_{a}+\dot{\omega}_{0}\Pi^{0}+\dot{\omega}B+\dot{P}\overline{c}+\dot{P}^{a}\overline{c}_{a}+\dot{c}\overline{P}+\dot{c}^{a}\overline{P}_{a}-\mathcal{H}_{U},\label{Q07}
\end{equation}
and the transition amplitude is given by
\begin{equation}
Z=\int De_{\mu}^{a}D\omega_{\mu}D\pi_{a}^{0}DB_{a}D\Pi^{0}DBD\eta^{A}D\overline{\mathcal{P}}_{A}\:\times\exp i\int dt\:\mathcal{L}_{U},\label{Q08}
\end{equation}

By construction, this amplitude must be independent of the choice of the gauge-fixing fermion. Usually, the fermion is given by $\Psi=\overline{\mathcal{P}}_{A}\chi^{A}$, where $\chi^{A}$ are gauge-fixing functions. Let us take $\chi^{A}=\left(\chi,\chi^{a},0,0\right)$, from where we have: $\Psi=\overline{c}\chi+\overline{c}_{a}\chi^{a}$.

In the temporal gauge, we choose $e_{0}^{a}=\delta_{1}^{a}$ and $\omega_{0}=0$. This gauge was introduced in the constraint analysis of two-dimensional $R^{2}$ gravity with torsion \cite{Kummer}, and in the context of Generalized Dilaton theories in \cite{Kummer-Dilaton}. The gauge-fixing functions can be written as
\begin{equation}
\chi=\frac{1}{\gamma}\omega_{0},\;\chi^{a}=\frac{1}{\gamma}\left(e_{0}^{a}-\delta_{1}^{a}\right),\label{Q09}
\end{equation}
where $\gamma$ is an arbitrary parameter, which is introduced due to the not dependence on the choice of gauge-fixing function in the transition amplitude. Therefore, in the temporal gauge we have
\begin{equation}
\left\{ \Psi,\Omega\right\} =-\frac{1}{\gamma}\omega_{0}\Pi^{0}-\frac{1}{\gamma}\left(e_{0}^{a}-\delta_{1}^{a}\right)\pi_{a}^{0}-\frac{1}{\gamma}P\overline{c}-\frac{1}{\gamma}P^{a}\overline{c}_{a}.\label{Q10}
\end{equation}

We can perform the displacement in the momenta $\Pi^{0}\rightarrow\gamma\Pi^{0}$, $\pi_{a}^{0}\rightarrow\gamma\pi_{a}^{0}$, and for the ghost fields $\overline{c}\rightarrow\gamma\overline{c}$, $\overline{c}_{a}\rightarrow\gamma\overline{c}_{a}$. This transformation has jacobian equal to one and does not change the measure of the integration. Notwithstanding, we can take the limit $\gamma\rightarrow 0$, once that the result must be independent of this choice.

After computing the integral in the non-dynamical fields: $\left(e_{0}^{a},\omega_{0}\right)$, and the fields $\left(P,\overline{P},P^{a},\overline{P}^{a}\right)$, we obtain
\begin{eqnarray}
Z & = & \int De_{1}^{a}D\omega DB_{a}DB\:\exp i\int dt\:\left[\dot{e}_{1}^{a}B_{a}+\dot{\omega}B+\frac{1}{l}\mathcal{G}_{1}\right]\times Z_{gh},\label{Q11a}\\
Z_{gh} & = & \int D\overline{c}DcD\overline{c}^{a}Dc^{a}\:\exp i\int dt\:\left[\overline{c}\dot{c}+l\overline{c}_{a}\dot{c}^{a}+\frac{1}{l}\epsilon_{a1}\left(1-4\kappa^{2}B\right)\overline{c}c^{a}-\epsilon_{\;1}^{a}\overline{c}_{a}c\right].\label{Q11b}
\end{eqnarray}
Notice that for the ghost sector, the is a coupling between the ghost the $B$ component of the Background field. This coupling is not present in the $BF$ action for Jackiw-Teitelboim gravity \cite{Chamseddine}.

\section{Quadratic Gravity from Dilaton Theory}

In the preceding sections, we studied two-dimensional quadratic gravity as a Polynomial $BF$ theory. Nonetheless, it can also be analyzed from the point of view of two-dimensional dilaton theories. Let us begin considering the following dilaton action
\begin{equation}
S\left(g,X\right)=\int d^{2}x\sqrt{g}\left[\frac{R}{2}X-V\left(X\right)\right],\label{Di01}
\end{equation}
where $R$ is the Ricci curvature scalar, $X$ the dilaton field, and $V=V\left(X\right)$ an arbitrary function. The action above represents a subclass of more general dilaton theories \cite{Grumiller} which have no degrees of freedom. Notwithstanding, it also corresponds to several cases of interest, for instance, if we set $V\left(X\right)=\Lambda X$, we obtain the Jackiw-Teitelboim action
\cite{Jackiw-Teitelboim}.

In general, from the action (\ref{Di01}) we obtain the equation of motion for the dilaton field
\begin{equation}
R=2\frac{dV}{dX}.\label{Di02}
\end{equation}
This equation relates the dilaton field $X$ with the scalar curvature $R$. Let us choose
\begin{equation}
V\left(X\right)=2\Lambda X\left(1-2\kappa^{2}X\right),\label{Di03}
\end{equation}
where $\kappa\neq 0$ is a constant parameter. Then, on-shell we have that the dilaton field and the potential are
now given by
\begin{equation}
X=\frac{1}{4\kappa^{2}}\left(1-\frac{R}{4\Lambda}\right),\: V\left(X\right)=\frac{\Lambda}{4\kappa^{2}}\left(1-\frac{R^{2}}{16\Lambda^{2}}\right).\label{Di04}
\end{equation}
Replacing these relations in the action (\ref{Di01}), we obtain the pure metric description of quadratic gravity (\ref{BF11}). Another quadratic potentials are also of great interest in dilaton theories: The KV model \cite{KV} is used to study $R^2$ gravity with torsion. The Almheiri-Polchinski model \cite{Almheiri} is also quadratic in the potential, however it has a different kinetic term. $R^2$ gravity subject to the constraint of constant curvature was studied in \cite{Odintsov}.

Instead of using the metric description, it is customary to work with a first-order action, which is defined in terms of the zweibein $e^{a}$, the spin-connection $\omega_{\; b}^{a}$, the dilaton $X$ and an auxiliary field $X_{a}$ as
\begin{eqnarray}
S & = & \int d^{2}x\left[Xd\omega+X_{a}\left(de^{a}+\omega_{\; b}^{a}\wedge e^{b}\right)-\frac{1}{2}\epsilon_{ab}e^{a}\wedge e^{b}V\left(X\right)\right].\label{Di05}
\end{eqnarray}

The variation along the auxiliary field $X_{a}$ gives as equation of motion the torsion-free condition. Therefore, whenever this condition is satisfied, we have $eR=2\varepsilon^{\mu\nu}\partial_{\mu}\omega_{\nu}$, and the first-order action (\ref{Di05}) becomes equivalent to (\ref{Di01}). In the specific case of quadratic gravity, we obtain
\begin{equation}
S=\int d^{2}x\left[X\left(\varepsilon^{\mu\nu}\partial_{\mu}\omega_{\nu}-2e\Lambda\right)+X_{a}\varepsilon^{\mu\nu}\left(\partial_{\mu}e_{\nu}^{a}+\epsilon_{\; b}^{a}\omega_{\mu}e_{\nu}^{b}\right)+4e\Lambda\kappa^{2}X^{2}\right].\label{Di06}
\end{equation}
This action is analog the one in (\ref{BF07}) if we identify the components of the Background field $B$ and $B_{a}$ as the dilaton $X=\frac{1}{2}\epsilon^{ab}B_{ab}$ and auxiliary field $X_{a}=B_{a2}$. The constraint analysis of the generalized dilaton action \cite{Kummer-Dilaton} shown that it has zero degrees of freedom, the same result obtained from the Polynomial $BF$ action. However, notice that, in the dilaton framework, quadratic gravity is obtained by choosing a given potential. In the $BF$ scheme, the Einstein-Hilbert and quadratic terms appears from an interaction term. Furthermore, while in Generalized dilaton theories, $X$ and $X_{a}$ are completely independent fields, in the Polynomial $BF$ action, they are independent components of the Background field $B_{IJ}$, which is initially $SO\left(3\right)$ invariant.

\section{Final Remarks}

In this work, we have dealt with the constraint analysis of the two-dimensional quadratic gravity. This theory is described by the Polynomial $BF$ action, which consists in a $SO\left(3\right)$ invariant $BF$ action plus a break invariance term. The parameter of this interaction is $\kappa^{2}$ and can be fixed by choosing the correct constant in front of the Einstein-Hilbert term. In case $\kappa^{2}=0$, we obtain the Jackiw-Teitelboim Gravity.

In contrast with the metric description of quadratic gravity, the Polynomial $BF$ action is of first-order in its derivatives. This fact lead us to a simply description of its constraint structure. The Dirac's programme closes at the second stage, i.e., after we found secondary constraints. These constraints are important once that they are the ones responsible for the gauge transformations of the dynamical variables. Furthermore, the algebra between the secondary constraints depends on the interaction factor $\kappa^{2}$ and the Background field component $B$.

Once that all constraints close an involutive algebra, we use the BFV path integral formulation. We introduced ghosts field $\left(P,P^{a},c,c^{a}\right)$ and their respective canonical conjugates for each constraint $\left(\phi,\phi_{a},\mathcal{G},\mathcal{G}_{a}\right)$. With these quantities we build of the BRST charge $\Omega$ and choose an appropriate gauge-fixing fermion. We computed the unitarized Hamiltonian and Lagrangian densities, which are necessary to build the transition amplitude in the Path Integral formalism.

We also briefly review the Generalized Dilaton theories. These theories are, undoubtedly, a powerful tool to study two-dimensional gravity, once that, given an appropriated potential for the dilaton field, we can obtain different gravity models, such as, Jackiw-Teitelboim or quadratic gravity. On the other hand, the Polynomial $BF$ action is build as a gauge theory with interaction, making this construction not only valid in two-dimensions but in higher dimensions as well. In addition, since the Polynomial $BF$ action is an extension of Jackiw-Teitelboim Gravity, there exists the possibility for a supersymmetric generalization, as shown in \cite{Chamseddine}, a non-commutative version, as in
\cite{nc-twodim}, or centrally extensions \cite{Centrally}, for instance.

As we mentioned, the two-dimensional Polynomial $BF$ action is closely related to the four-dimensional MacDowell-Mansouri Gravity, and our canonical analysis needs no gauge fixing condition at the classical level, in contrast with the analysis of the MacDowell-Mansouri $BF$ theory shown in \cite{Durka}. Therefore, our analysis can shed some light in the canonical analysis of the four dimensional $BF$ theory using the Hamilton-Jacobi formalism \cite{HJ}, for example. Furthermore, the relation between Polynomial BF and dilaton theory can provide insight in the construction of new interaction term, allowing the introduction of dynamical torsion, which can also be generalized in four dimensions.

\section{Acknowledgements}

I am grateful to R. Paszko, B. M. Pimentel and D. Vassilevich for reading the article and useful comments. C. E. Valc\'arcel was supported by FAPESP process $2012/23520-7$ and CNPq process $150407/2016-5$.

\section{Appendix: Conventions and Notation}

During the article we follow the notation of the textbook \cite{Ortin}. The $SO\left(n\right)$ is the group of matrices with determinant $1$ and preserve the metric $\eta_{IJ}=diag\left(+,+,...,+\right)$. The generators $M_{IJ}=-M_{JI}$ of its correspondent algebra satisfy the following relation
\begin{equation}
\left[M_{IJ},M_{KL}\right]=\eta_{IL}M_{JK}-\eta_{IK}M_{JL}+\eta_{JK}M_{IL}-\eta_{JL}M_{IK}.\label{AP01}
\end{equation}

For the group $SO\left(3\right)$ we consider the metric $\eta_{IJ}=diag\left(+,+,+\right)$, where $I=1,2,3$. We can decompose the generators $M_{IJ}$ into the generators $M_{ab}$ $\left(a=0,1\right)$ and the generators $M_{a2}$:
\begin{eqnarray*}
\left[M_{ab},M_{cd}\right] & = & \eta_{ad}M_{bc}-\eta_{ac}M_{bd}+\eta_{bc}M_{ad}-\eta_{bd}M_{ac},\\
\left[M_{ab},M_{c2}\right] & = & -\eta_{ac}M_{b2}+\eta_{bc}M_{a2},\\
\left[M_{a2},M_{b2}\right] & = & -M_{ab}.
\end{eqnarray*}
Since we are working in two dimensions, we can write $M_{ab}=\epsilon_{ab}M$ and, if we define $P_{a}\equiv M_{a2}$, it is possible to rewrite the algebra as:
\begin{eqnarray}
\left[M,M\right] & = & 0,\;\left[M,P_{a}\right]=-\epsilon_{a}^{\;\;b}
P_{b},\:\left[P_{a},P_{b}\right]=-\epsilon_{ab}M.\label{AP02}
\end{eqnarray}
Note that, for $SO\left(3\right)$, we can redefine the generators as: $J_{I}\equiv\frac{1}{2}\epsilon_{I}^{\;\; JK}M_{JK}$. However, the notation with generators $M_{IJ}$ can be used in the three-dimensional $BF$ Polynomial action as well. Furthermore, by using a Minkowski metric $\eta_{IJ}$, we can generalize the formalism to the group $SO\left(2,1\right)$ an explore the De Sitter or Anti-De Sitter spaces.

The pure $BF$ model under the group $SO\left(3\right)$ is given by
\begin{eqnarray}
S_{2D} & = & -\frac{1}{2}\int\: tr\left(B\wedge F\right)=\frac{1}{4}\int d^{2}x\:\varepsilon^{\mu\nu}B_{IJ}F_{\mu\nu}^{IJ},\label{AP03}
\end{eqnarray}
where the components of the field strength are
\begin{equation}
F_{\mu\nu}^{IJ}=\partial_{\mu}A_{\nu}^{IJ}-\partial_{\nu}A_{\mu}^{IJ}+A_{\mu K}^{I}A_{\nu}^{KJ}-A_{\nu K}^{J}A_{\mu}^{KI}.\label{AP04}
\end{equation}
Here we have computed the trace of (\ref{AP03}) in the adjoint representation: $tr\left[M_{IJ}M_{KL}\right]=-2\left(\eta_{IK}\eta_{JL}-\eta_{IL}\eta_{JK}\right)$. The equations of motion are given by
$F=0$ and $DB=0$, where the explicit form of this covariant derivative is
\begin{equation}
D_{\mu}X_{IJ}\equiv\partial_{\mu}X_{IJ}+A_{I\mu}^{\;\; K}X_{KJ}-A_{J\mu}^{\;\; K}X_{KI}.\label{AP05}
\end{equation}
By decomposing the gauge and background fields in the Lorentz and translation components, this covariant derivative becomes
\begin{eqnarray}
D_{\mu}X_{a} & = & \partial_{\mu}X_{a}+\epsilon_{a}^{\;\; c}X_{c}\omega_{\mu}-\frac{1}{l}\epsilon_{ac}e_{\mu}^{c}X.\label{AP06a}\\
D_{\mu}X & = & \partial_{\mu}X+\frac{1}{l}\epsilon_{\;\; b}^{a}e_{\mu}^{b}X_{a}.\label{AP06b}
\end{eqnarray}

The pure $BF$ model (\ref{AP03}) is invariant under the gauge transformation
\begin{equation}
\delta A_{\mu}^{I}=D_{\mu}\xi_{IJ},\:\delta B_{IJ}=-\xi_{I}^{\;\; K}B_{KJ}+\xi_{J}^{\;\; K}B_{KI}.\label{AP07}
\end{equation}


\bigskip

\begin{thebibliography}{99}

\bibitem{MacDowell}S. W. MacDowell, F. Mansouri, \textit{Unified Geometric Theory of Gravity and Supergravity}, Phys.
Rev. Lett. $\boldsymbol{38}$, $(1977)$, 739.

\bibitem{Olea}R. Aros, M. Contreras, R. Olea, R. Troncoso, J. Zanelli, \textit{Conserved charges for gravity with
locally AdS asymptotics}, Phys. Rev. Lett. $\boldsymbol{84}$, $(2000)$, 1647;\\
O. Miskovic, R. Olea, \textit{Topological regularization and self-duality in four-dimensional anti-de Sitter gravity}, Phys. Rev. D $\boldsymbol{79}$, $(2009)$, 124020.

\bibitem{Smolin} L. Smolin, \textit{Holographic formulation of quantum general relativity}, Phys. Rev.
D $\boldsymbol{61}$, $(2000)$, 084007;\\
L. Smolin, A. Starodubtsev, \textit{General relativity with a topological phase: an action principle}, arXiv:hep-th/0311163, $(2003)$.

\bibitem{Freidel} L. Freidel, A. Starodubtsev, \textit{Quantum gravity in terms of topological
observables}, arXiv:hep-th/0501191, $(2005)$;\\
L. Freidel, S. Speziale, \textit{On the Relations between Gravity and BF Theories}, SIGMA $\boldsymbol{8}$, $(2012)$, 032.

\bibitem{Perez} A. Perez, \textit{The spin foam approach to Quantum Gravity}, Living Rev. Rel.
$\boldsymbol{16}$, $(2013)$, 3.

\bibitem{Ashtekar} A. Ashtekar, J. Lewandowski, \textit{Background independent quantum gravity: A status report}, Class.
Quantum Grav. $\boldsymbol{21}$, $(2004)$, R53.

\bibitem{Brown} J. D. Brown, \textit{Lower Dimensional Gravity}, World Scientific Pub. Co. Inc. $(1988)$.

\bibitem{Jackiw-Teitelboim} C. Teitelboim, \textit{Gravitation and Hamiltonian Structure in Two Space-Time
Dimensions}, Phys. Lett. B $\boldsymbol{126}$, $(1983)$, 41;\\
$\boldsymbol{126}$,
C. Teitelboim, \textit{Supergravity And Hamiltonian Structure In Two Space-time Dimensions}, Phys. Lett. B $\boldsymbol{126}$, $(1983)$, 46;\\
R. Jackiw, C. Teitelboim, \textit{Quantum Theory of Gravity}, edited by S. Christensen, Adam Hilger, Bristol, $(1984)$;\\
R. Jackiw, \textit{Lower Dimensional Gravity}, Nucl. Phys. B $\boldsymbol{252}$, $(1985)$ 343.

\bibitem{Grumiller} S. Nojiri, S. D. Odintsov, \textit{Quantum dilatonic gravity in $d=2$, $4$ and $5$
dimensions}, Int. J. Mod. Phys. A $\boldsymbol{16}$, $(2001)$ 1015;\\
D. Grumiller, W. Kummer, D. V. Vassilevich, \textit{Dilaton gravity in two dimensions},
Phys. Rep. $\boldsymbol{369}$, $(2002)$ 327.

\bibitem{Isler} K. Isler, C. A. Trugenberger, \textit{Gauge theory of Two-Dimensional Quantum Gravity},
Phys. Rev. Lett $\boldsymbol{63}$, $(1989)$ 834.

\bibitem{Chamseddine} A. H. Chamseddine, D. Wyler, \textit{Gauge theory of Topological Gravity in $1+1$ dimensions},
Phys. Lett. B $\boldsymbol{228}$, $(1989)$, 75;\\
A. H. Chamseddine, D. Wyler, \textit{Topological Gravity in $1+1$ dimensions}, Nucl. Phys. B $\boldsymbol{340}$, $(1990)$, 595.

\bibitem{Paszko} R. Paszko, R. da Rocha, \textit{Quadratic gravity from BF theory in two and three dimensions}, Gen.
Rel. Grav. $\boldsymbol{47}$, $(2015)$ 94.

\bibitem{Felice} A. De Felice, S. Tsujikawa, \textit{$f(R)$ theories}, Living Rev. Rel. $\boldsymbol{13}$,
$(2010)$ 3;\\
S. Nojiri, S. D. Odintsov, \textit{Unified cosmic history in modified gravity: from $F(R)$ theory to Lorentz non-invariant models}, Phys. Rept. $\boldsymbol{505}$, $(2011)$ 59.

\bibitem{Dirac} P. A. M. Dirac, \textit{Generalized Hamiltonian dynamics}, Can. J. Math. $\boldsymbol{2}$,
$(1950)$, 129;\\
P. A. M. Dirac, \textit{The Hamiltonian form of field dynamics}, Can. J. Math. $\boldsymbol{3}$, $(1951)$ 1;\\
P. A. M. Dirac, \textit{Lectures on Quantum Mechanics}, New York: Yeshiva University, $(1964)$.

\bibitem{Piguet} C. P. Constantinidis, J. A. Louren\c{c}o, I. Morales, O. Piguet, A. Rios, \textit{Canonical Analysis
of the Jackiw-Teitelboim Model in the Temporal Gauge.  I. The Classical Theory}, Class. Quant.
Grav. $\boldsymbol{25}$, $(2008)$ 125003;\\
C. P. Constantinidis, O. Piguet, A. Perez, \textit{Quantization of the Jackiw-Teitelboim model}, Phys.
Rev. D $\boldsymbol{79}$, $(2009)$, 084007.

\bibitem{BF-HJ} M. C. Bertin, B. M. Pimentel, C. E. Valc\'arcel, \textit{Two-dimensional background field gravity:
A Hamilton-Jacobi analysis}, J. Math. Phys. $\boldsymbol{53}$, $(2012)$ 102901.

\bibitem{HJ}  Y. G\"uler, \textit{On the dynamics of singular, continuous systems}, J. Math. Phys.
$\boldsymbol{30}$, $(1989)$ 785;\\
M. C. Bertin, B. M. Pimentel, C. E. Valc\'arcel, \textit{Non-Involutive Constrained Systems and Hamilton-Jacobi
Formalism }, Annals Phys. $\boldsymbol{323}$, $(2008)$ 3137;\\
M. C. Bertin, B. M. Pimentel, C. E. Valc\'arcel, \textit{Involutive constrained systems and Hamilton-Jacobi formalism}, J. Math. Phys. $\boldsymbol{55}$, $(2014)$ 112901.

\bibitem{BFV} E. S. Fradkin, G. A. Vilkovisky, \textit{Quantization Of Relativistic Systems With Constraints},
Phys. Lett. B $\boldsymbol{55}$, $(1975)$ 244; \\
A. Batalin, G. A. Vilkovisky, \textit{Relativistic S Matrix of Dynamical Systems with Boson and Fermion Constraints}, Phys. Lett. B $\boldsymbol{69}$ $(1977)$ 309;\\
E. S. Fradkin, T. E. Fradkina, \textit{Quantization of Relativistic Systems with Boson and Fermion First and Second Class Constraints}, Phys. Lett. B $\boldsymbol{72}$, $(1978)$ 343.

\bibitem{Henneaux} M. Henneaux, C. Teitelboim, \textit{Quantization of Gauge Systems}, Princeton Univ.
Press $(1992)$.

\bibitem{Rothe} H. J. Rothe, K. D. Rothe, \textit{Classical and Quantum Dynamics of Constrained Hamiltonian
Systems}, World Scientific $\boldsymbol{81}$, $(2010)$.

\bibitem{Gambini} R. Gambini, J. Pullin, \textit{Loops, Knots, Gauge Theories and Quantum Gravity}, Cambridge
University Press, $(2000)$.

\bibitem{Kummer} W. Kummer, D. J. Schwarz, \textit{Renormalization of $R^{2}$ gravity with dynamical torsion in
$d=2$}, Nucl. Phys. B $\boldsymbol{382}$, $(1992)$, 171.

\bibitem{Kummer-Dilaton} W. Kummer, H. Liebl, D.V. Vassilevich, \textit{Exact Path Integral Quantization of
Generic $2-D$ Dilaton Gravity}, Nucl. Phys. B $\boldsymbol{493}$, $(1997)$, 491.

\bibitem{KV} M. O. Katanayev, I. V. Volovich, \textit{String Model with Dynamical Geometry and Torsion}, Phys. Lett.
B $\boldsymbol{175}$, (1986), 413;\\
M. O. Katanaev, I. V. Volovich, \textit{Two-dimensional Gravity with Dynamical Torsion and Strings}, Ann. Phys.
$\boldsymbol{197}$, (1990), 1.

\bibitem{Almheiri} A. Almheiri, J. Polchinski, \textit{Models of $AdS2$ backreaction  and  holography}, JHEP
$\boldsymbol{1511}$ (2015) 014.

\bibitem{Odintsov} T. Muta, S. D. Odintsov, \textit{Two-dimensional higher derivative quantum gravity with
constant curvature constraint}, Prog.Theor.Phys. $\boldsymbol{90}$, $(1993)$ 247; Phys. Atom. Nucl. $\boldsymbol{56}$,  $(1993)$ 1121; Yad. Fiz. $\boldsymbol{56}$, $(1993)$ no.8, 223.

\bibitem{nc-twodim} S. Cacciatori, A. H. Chamseddine, D. Klemm, L. Martucci, W.A. Sabra, D.
Zanon, \textit{NoncommutativeGravity in Two dimensions}, Class. Quantum Grav. $\boldsymbol{19}$, $(2002)$ 4029;\\
D. V. Vassilevich, \textit{Quantum noncommutative gravity in two dimensions}, Nuclear Physics B $\boldsymbol{715}$, $(2005)$ 695.

\bibitem{Centrally} H. Verlinde, \textit{Black Holes and Strings in Two Dimensions}, proceeding published in
\textit{Trieste Spring School on Strings and Quantum Gravity}, $(1991)$, 178;\\
D. Cangemi, R. Jackiw, \textit{Gauge Invariant Formulations of Linear Gravity}, Phys. Rev. Lett. $\boldsymbol{69}$, $(1992)$, 233.

\bibitem{Durka} R. Durka, J. Kowalski-Glikman, \textit{Hamiltonian analysis of $SO(4,1)$ constrained BF
theory}, Class.Quant.Grav. $\boldsymbol{27}$, $(2010)$ 185008.

\bibitem{Ortin} T. Ortin, \textit{Gravity and Strings}, Cambridge University Press, $(2004)$.

\end{thebibliography}
\end{document}